\documentclass{llncs}
\usepackage{graphicx}
\usepackage{caption,flushend}
\usepackage{colortbl}
\usepackage{booktabs}
\definecolor{mygreen}{rgb}{0.7,1,0.7}
\definecolor{myblue}{rgb}{0,0.9,1}
\definecolor{mygray}{rgb}{0.8,0.8,0.8}

\begin{document}

\title{Aggregation Schemes for Single-Vector WSI Representation Learning in Digital Pathology}

\author{-} 
\author{\emph{Sobhan Hemati, Ghazal Alabtah, Saghir Alfasly, H.R. Tizhoosh}}
\institute{KIMIA Lab, Department of Artificial Intelligence and Informatics, \\Mayo Clinic, Rochester, MN, USA} 
\maketitle
\begin{abstract}
A crucial step to efficiently integrate Whole Slide Images (WSIs) in computational pathology is assigning a single high-quality feature vector, i.e., one embedding, to each WSI. With the existence of many pre-trained deep neural networks and the emergence of foundation models, extracting embeddings for sub-images (i.e., tiles or patches) is straightforward. However, for WSIs, given their high resolution and gigapixel nature, inputting them into existing GPUs as a single image is not feasible. As a result, WSIs are usually split into many patches. Feeding each patch to a pre-trained model, each WSI can then be represented by a set of patches, hence, a set of embeddings. Hence, in such a setup, WSI representation learning reduces to set representation learning where for each WSI we have access to a set of patch embeddings. To obtain a single embedding from a set of patch embeddings for each WSI, multiple set-based learning schemes have been proposed in the literature. In this paper, we evaluate the WSI search performance of multiple recently developed aggregation techniques (mainly set representation learning techniques) including simple average or max pooling operations, Deep Sets, Memory networks, Focal attention, Gaussian Mixture Model (GMM) Fisher Vector, and deep sparse and binary Fisher Vector on four different primary sites including bladder, breast, kidney, and Colon from TCGA. Further, we benchmark the search performance of these methods against the median of minimum distances of patch embeddings, a non-aggregating approach used for WSI retrieval. 
\end{abstract}

\section{Introduction}

The image analysis field has witnessed remarkable advancements in recent years, thanks to the integration of deep learning models and the availability of large-scale data. However, analysis of histopathology images using deep learning is not as straightforward as ordinary-sized images. Especially, to build whole slide image (WSI) search engines, the most critical step is to be able to measure the similarity between WSIs efficiently and accurately \cite{tizhoosh2024image,Lahr2024}. The state-of-the-art approach to this step is to extract high-quality deep embeddings that capture tissue morphology from each WSI using a well-trained deep model. To this end, it is necessary to feed WSIs to build such models. However, given the gigapixel nature of WSIs, feeding a WSI at high magnification into a model to GPU memory is infeasible. To overcome this challenge, a WSI is usually split into a set of much smaller sub-images called ``patches'' or tiles. Following this procedure, we end up with a set of deep embeddings per WSI which makes it challenging to employ them for downstream tasks like WSI retrieval (a set of pacthes that could be very large if a smart patch selection is not available). In this situation, having one single-vector WSI embedding can mitigate such challenges. With single-vector WSI representation  much less memory is required to build a WSI search system. As well, using distance measures calculating similarity between WSIs becomes trivial. Finally, the search speed will increased and storage requirements will decrease resulting in wider deployment of WSI image classification and retrieval.

\begin{figure}
    \centering
    \includegraphics[width=\linewidth]{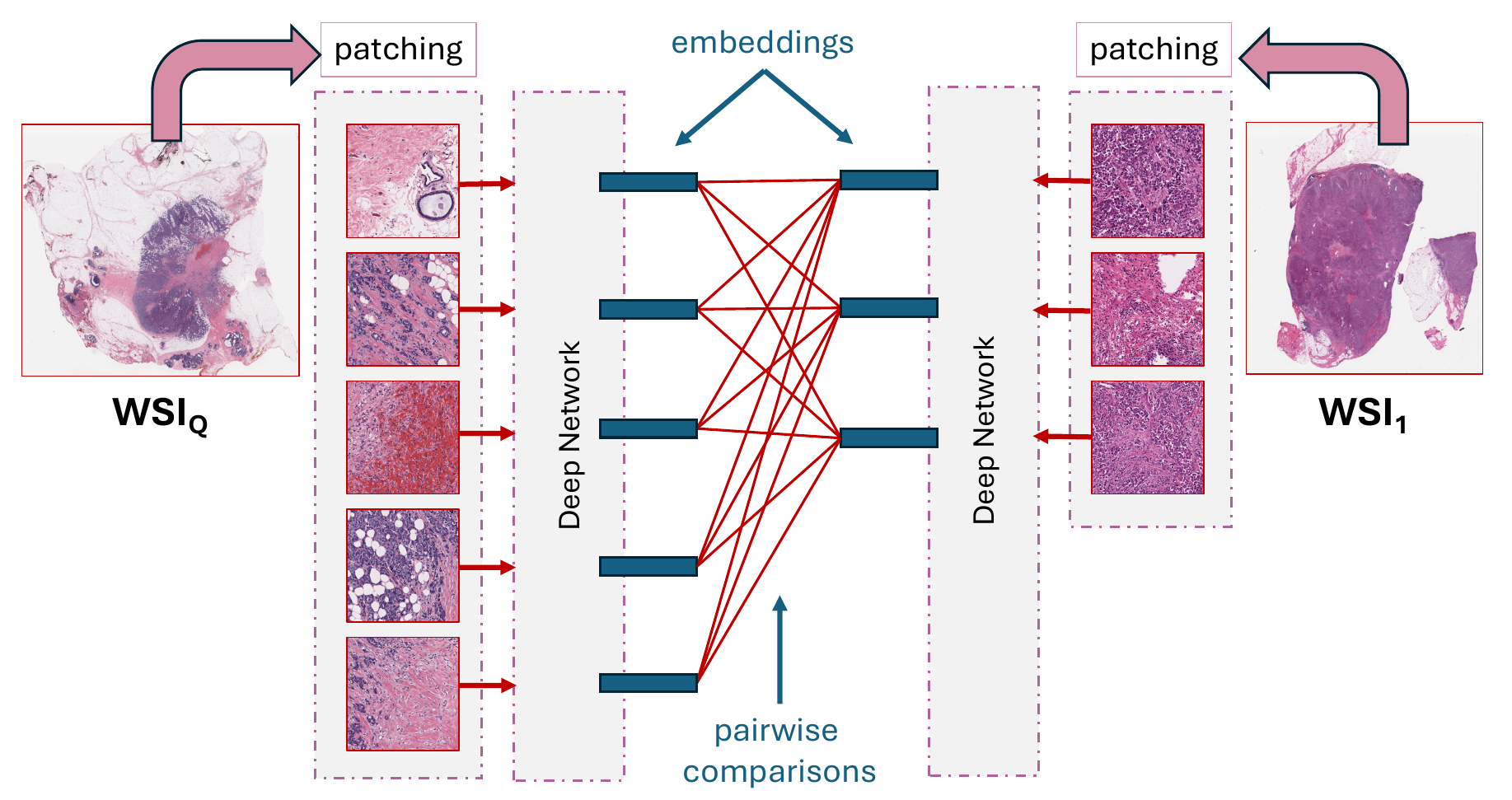}
    \caption{Yottixel \cite{kalra2020yottixel}, although a patch-oriented search engine, introduced the first practical approach for WSI-to-WSI comparison through the median-of-minimum distance measurements.}
    \label{fig:enter-label}
\end{figure}

\begin{figure}
    \centering
    \includegraphics[width=\linewidth]{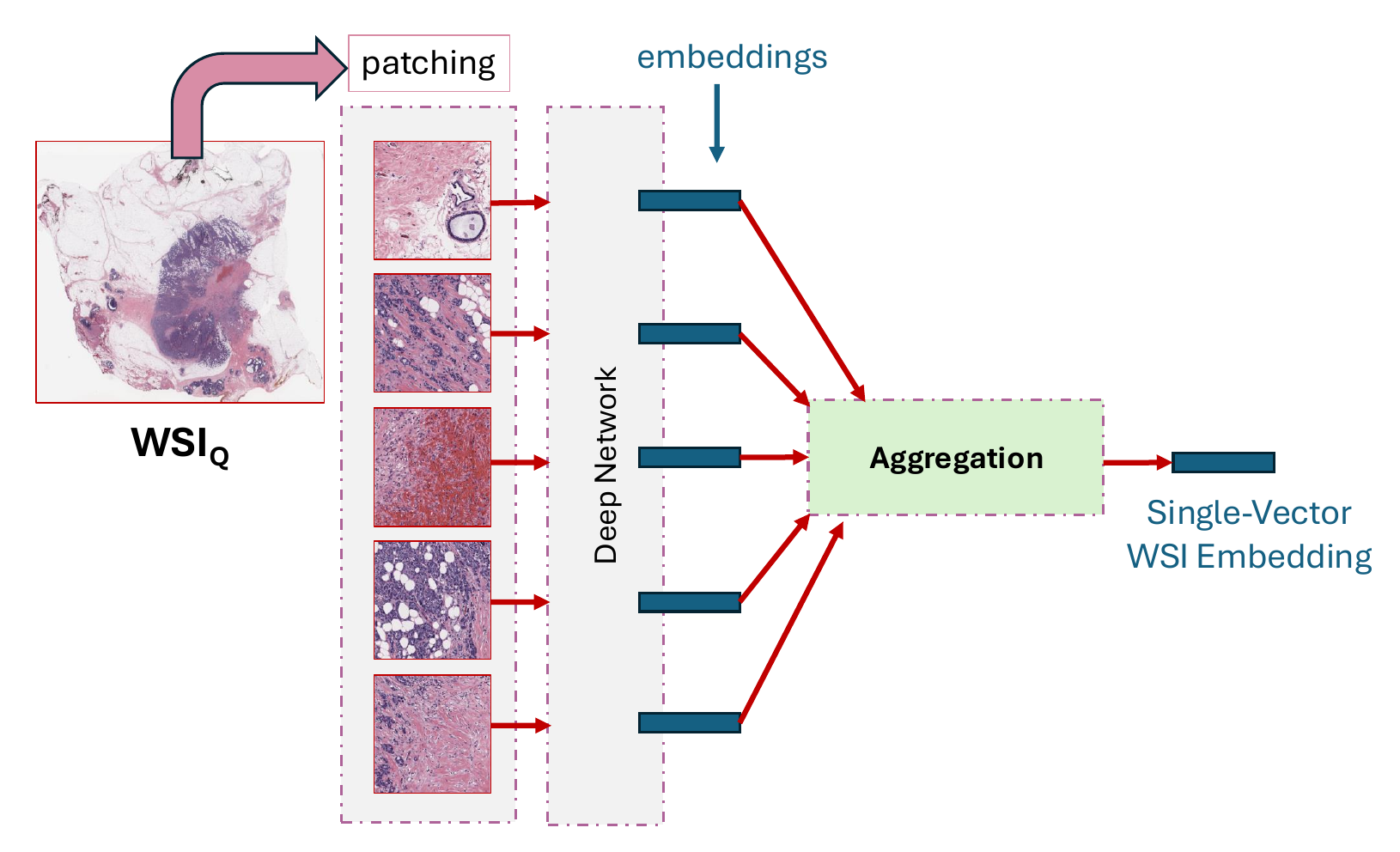}
    \caption{Aggregation of patch embeddings can provide a single-vector WSI representation.}
    \label{fig:singlevectoraggreg}
\end{figure}

A practical and established approach to calculate the similarity between two sets of embeddings is the \emph{median-of-minimums}, a scheme that was introduced as part of the Yottixel search engine \cite{kalra2020yottixel}; one calculates the minimum distance of any given input patch to all other patches (of the other WSI) and takes the median of all minimum distances overall (see Fig. \ref{fig:enter-label}). This approach is quite practical and fast (as Yottixel used binary feature vectors) but it is not able to provide a single vector to represent WSI. A more desirable solution would be to obtain \emph{one single embedding} per WSI for more efficient storage and more targeted WSI-to-WSI matching, or for WSI classification (Fig. \ref{fig:singlevectoraggreg}). 

To this end, different aggregation algorithms to derive a single vector of deep features from a set of patch embeddings have been proposed including simple average or max pooling operations, Deep Sets \cite{zaheer2017deep}, memory networks \cite{kalra2020learning}, focal attention \cite{kalra2021pay}, graph convolution neural networks \cite{adnan2020representation}, Gaussian mixture model (GMM) Fisher Vector \cite{jaakkola1999exploiting,perronnin2007fisher,hemati2023learning}, and Deep Fisher Vector and its variations (binary and sparse Fisher Vector) \cite{hemati2023learning}.

Although multiple set representation learning techniques have been proposed and investigated in the literature, to the best of our knowledge, there is no study that evaluates these algorithms against each other for WSI retrieval as a challenging task. To address this shortcoming, in this paper, we briefly introduce each set representation learning technique and discuss our benchmarking scheme. Subsequently, we evaluate each method over four different datasets namely bladder, breast, kidney, and colon through the $k$-Nearest Neighbour ($k$-NN) search using WSI embeddings obtained from each method. Finally, in the result section we present and discuss the performance of each technique to generate a single vector of deep features for a WSI.

\subsection{Statement of Significance}
Whole Slide Image (WSI) representation learning is a cornerstone of computational pathology,
enabling efficient retrieval, classification, and analysis of gigapixel-scale histopathology images.
However, the challenge lies in deriving a single high-quality feature vector for an entire WSI,
given that existing GPUs cannot process them as single images due to their enormous resolution.
Traditionally, WSIs are divided into smaller patches, each independently embedded using deep
neural networks, leading to a set-based representation problem where optimal aggregation
strategies remain an open question.
This study systematically evaluates multiple state-of-the-art set representation learning
techniques—including Deep Sets, Memory Networks, Focal Attention, Gaussian Mixture Model
(GMM) Fisher Vector, and deep sparse and binary Fisher Vector—against conventional pooling
methods for WSI embedding aggregation. Additionally, we benchmark these approaches against
a non-aggregating retrieval technique based on the median of minimum distances of patch
embeddings. Our comprehensive evaluation, conducted on TCGA datasets across four cancer
types (bladder, breast, kidney, and colon), provides critical insights into the effectiveness of
different aggregation strategies for WSI search and retrieval.
By rigorously comparing various WSI embedding aggregation techniques, this study advances
the field of computational pathology by guiding researchers and practitioners toward more
effective and computationally feasible WSI retrieval methodologies. The findings have
significant implications for large-scale histopathology archives, image-based precision medicine,
and AI-driven diagnostic workflows.

\section{Background}
In this section, we briefly review set representation learning algorithms applied to the WSI representation learning problem.

\subsection{GMM Fisher Vector (1999-2007)} Fisher Vector can be seen as an extension to bag of visual words (BoVW) \cite{csurka2004visual} that was initially developed in computer vision research for encoding a set of local image descriptors into one embedding. Theoretically, Fisher Vector can be developed on top of any generative models \cite{jaakkola1999exploiting}. Some works \cite{perronnin2007fisher} have introduced  Gaussian mixture model (GMM)-based Fisher Vector which can be calculated using the gradient of the log-likelihood of the GMMs with respect to its parameters  given a bag of data points (e.g., means, variances, and mixing coefficients). Theoretically, while BoVW only employs mean statistics to obtain the set representation, in the GMM-based Fisher Vector, the variance as a higher-order statistic is also used.

\subsection{Deep Sets (2017-2021)} 
The Deep Set approach \cite{zaheer2017deep} focuses on permutation invariant representation learning and establishes a standard definition of the permutation invariant functions. This approach shows such a family of functions can be implemented as neural networks to learn permutation invariant representations. Other works employ Deep Sets-like architecture along with CNN to learn end-to-end WSI representations in WSI search tasks \cite{hemati2021cnn}. These methods showed the obtained representations achieve better search performance both in terms of search speed and accuracy compared with Yottixel search engine \cite{kalra2020yottixel}. 

\subsection{Memory Networks (2020)} 
Other works have presented Memory-based Exchangeable Model (MEM) to learn the permutation invariant set functions   \cite{kalra2020learning}. In this method, a MEM is constructed from multiple memory blocks where each memory block is a sequence-to-sequence model containing multiple memory units. The output of each memory block is invariant to the permutations of the input sequence. The memory units are in charge of calculation of attention values for each instance. To calculate these attention values, the memory vectors are aggregated using a pooling operation (weighted averaging) to form a permutation-invariant representation. Increasing the number of memory units enables the memory block to capture more complex dependencies between the instances of a set by providing an explicit memory representation for each instance in the sequence. MEM was evaluated on point cloud and lung WSIs classification.

\subsection{Focal Attention (2021)}
Another set representation scheme developed for WSI classification/search is Focal attention   \cite{kalra2021pay}. This approach was inspired by two recent developments in representation learning literature, focal loss and attention based MIL \cite{ilse2018attention} and proposed the novel pooling layer called focal attention (FocAtt-MIL) where all patches of a mosaic are mapped into a single embedding using an attention-weighted averaging  \cite{ilse2018attention} modulated by a trainable focal factor. More precisely, FocAtt-MIL is composed from four main components. Prediction MLP, WSI Context, Attention Module, and Focal Network. The Prediction MLP is a trainable neural network that calculates a prediction for each patch embedding in the set. WSI context is also a neural network (basically Deep Sets from \cite{zaheer2017deep} that obtains one embedding per WSI in a simple efficient manner that capture general information from the WSI. The attention module is made from two MLPs transformation, and the Attention networks. The attention module takes the patch embedding and WSI context and output the an attention value between 0 to 1 for each instance. Finally, the focal network receive the WSI context and computes a focal factor per dimension which further guides the final prediction toward better WSI representation.

\subsection{Deep Fisher Vector and its variations (2023)}
\begin{figure}[htb]
  \centering
  \includegraphics[width=\linewidth]{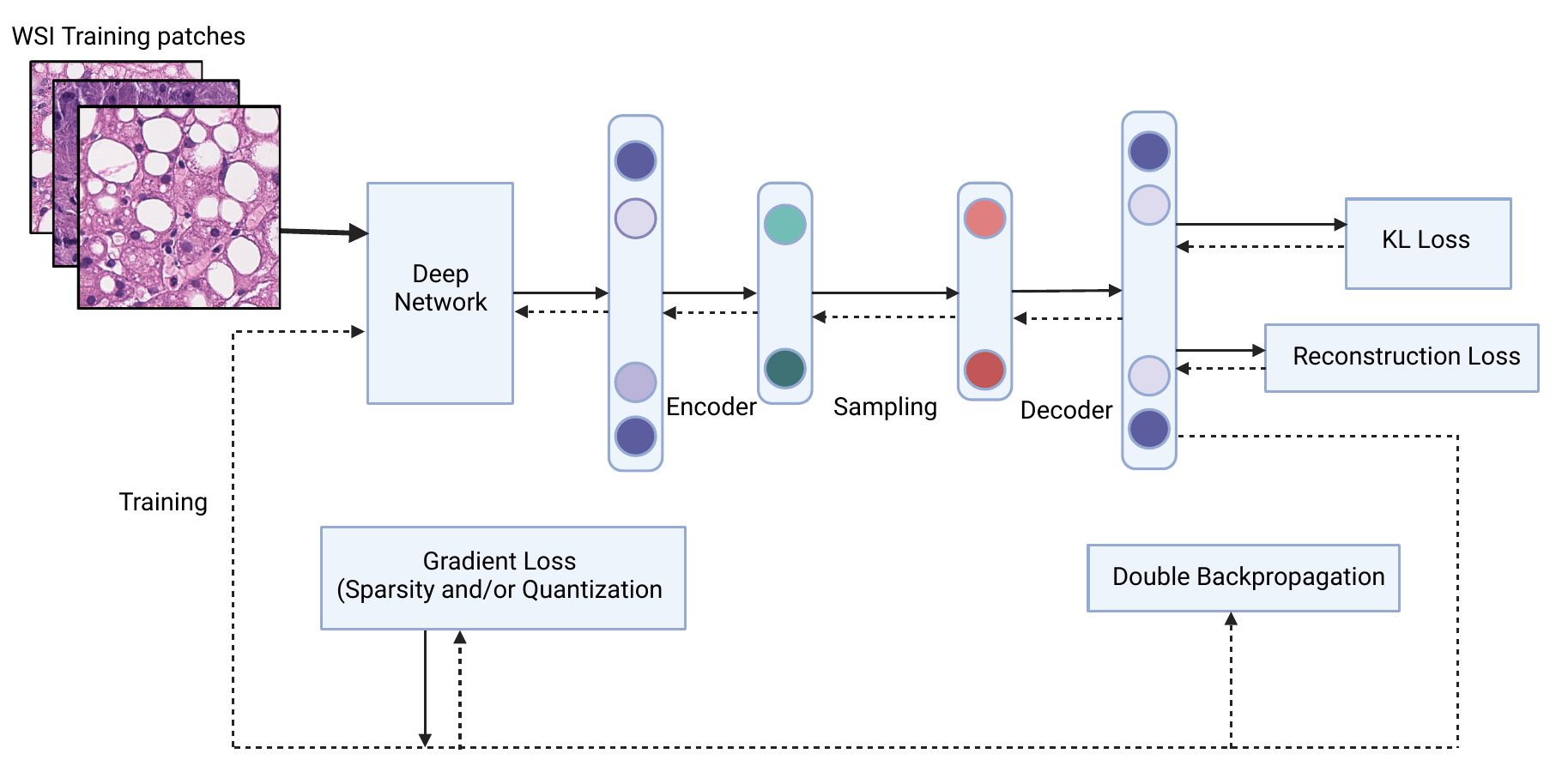}
  \caption{VAE Training architecture}
  \label{fig:diagram1_above}
\end{figure}

Motivated by the success of GMM-based Fisher Vector, Fisher Vector theory has been proposed for more recent deep generative models including Variational Autoencoders (VAEs) \cite{qiu2017deep} and GANs \cite{zhai2019adversarial}. Recently, other works have extended VAE-based Fisher Vectors for the WSI representation learning task  \cite{hemati2023learning}. More specifically, this approach offered two variations of Deep Fisher Vectors,  namely ``\emph{deep sparse Fisher Vector}'' and ``\emph{deep binary Fisher Vector}'',  to obtain binary and sparse permutation-invariant WSI embeddings suitable for downstream tasks such as efficient WSI search. To obtain the deep sparse, and binary Fisher Vectors, one can employ a simple VAE where its latent space is also connected to a dense layer with a softmax activation function that outputs a probability distribution representing the predicted WSI class. This classifier injects class information (i.e., diagnosis) into the final WSI embeddings. Considering that the Fisher Vector is derived from the gradient space of the VAE, to achieve sparse or binary embeddings, it is desired to have sparse or minimum quantization loss gradients. This step has been performed with a gradient regularization loss which is calculated using double backpropagation \cite{hemati2023learning}. Finally, the total loss for training the VAE consists of a weighted summation of reconstruction loss, KL divergence loss, classification loss, and gradient regularization loss. The contribution of gradient regularization loss to the total loss is controlled through the $\alpha$ regularization parameter. Fig. \ref{fig:diagram1_above} shows the training architecture.

\begin{figure}[htb]
  \includegraphics[width=\linewidth]{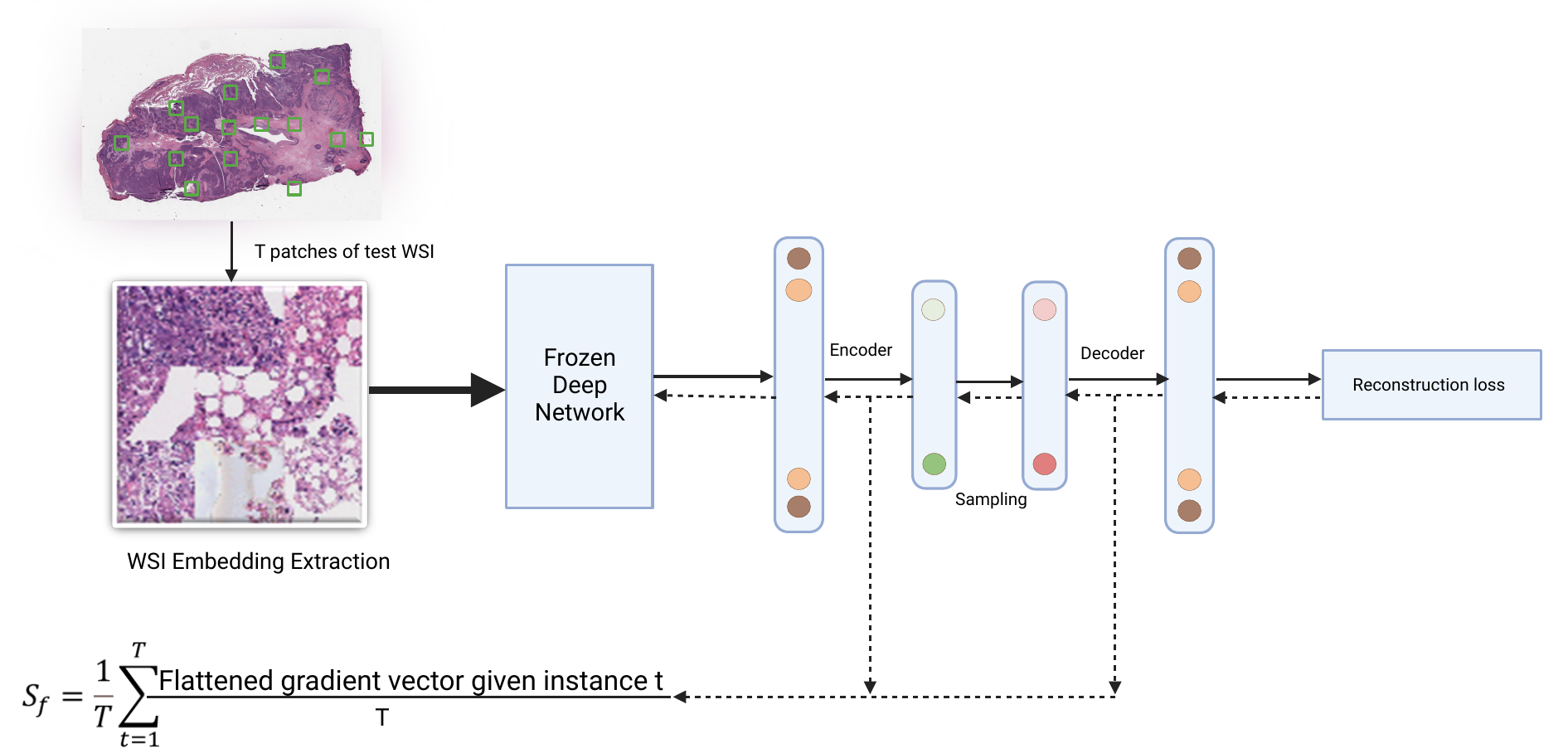}
  \caption{Aggregation to generate a single-vector WSI embedding.}
  \label{fig:diagram2_below}
\end{figure}

After training the VAE, first, the gradients of the reconstruction loss are computed. Second, the gradient using power and $L_2$ normalization steps are normalized (see \cite{perronnin2010improving}). Finally, considering that the dimensionality of the deep Fisher Vector is the same as the number of the VAE parameters (which can be a large number making the WSI search expensive), the effective dimensions are selected to be the top $M$ parameters that provide the highest variance in their respective gradient values for the training data. Fig. \ref{fig:diagram2_below} shows the calculation of single-vector WSI representation given the trained VAE and patch embeddings for a WSI. 

\subsection{Other Approaches}
Recently foundation models for multiple WSI representation learning have been proposed that use various techniques for aggregation of patch embedding, including GIGAPATH \cite{xu2024whole}, PRISM \cite{shaikovski2024prism}, CHIEF \cite{wang2024pathology}, and THREADS \cite{vaidya2025molecular}. Among these works, GIGAPATH \cite{xu2024whole} and THREADS \cite{vaidya2025molecular} require embeddings for all patches from WSI to perform the aggregation. Given that the backbones are generally foundation models, such techniques are extremely expensive in practice. Additionally, CHIEF \cite{wang2024pathology}  and PRISM \cite{shaikovski2024prism} use techniques based on the attention-based MIL by \cite{ilse2018attention}. Considering that here we already discussed two different extensions of the attention-based MIL namely Focal attention   \cite{kalra2021pay} and Memory-based Exchangeable Model   \cite{kalra2020learning} techniques, we did not include the aggregation techniques in CHIEF \cite{wang2024pathology}  and PRISM.

In terms of acquiring expressive embeddings, foundation models have been reported to be inaccurate for zero-shot WSI retrieval \cite{alfasly2025validation}, insensitive to inter-center variations \cite{de2025current}, and exhibit low rotational invariance \cite{elphick2024latent}. Moreover, they consume approximately 35 times more energy than conventional deep models, are not easily fine-tunable, and still require large amounts of data for downstream tasks \cite{mulliqi2025foundation}.   
\section{Methodology}
This paper aims to compare the performance of the multiple set aggregation techniques for WSI representation learning including simple average or max pooling operations, Deep Sets, Memory networks, Focal attention, Gaussian Mixture Model (GMM) Fisher Vector, and deep sparse and binary Fisher Vector. We validate these methods for the task of WSI retrieval. Hence, we must construct an evaluation framework involving different datasets, and baselines. In this section, first, we discuss the datasets, the benchmarking task and baselines.

\subsection{Datasets}

The datasets presented in this study contain a collection of four histopathology cases, including bladder, breast, kidney, and colon primary sites disease cases. All datasets are drawn from The Cancer Genome Atlas (TCGA). For each WSI, multiple patches were extracted using  Yottixel's mosaic algorithm (a two-staged clustering approach that employs stain histograms and patch proximity to group patches) \cite{kalra2020yottixel}. Each patch is represented by an embedding extracted from DenseNet \cite{huang2017densely} with imageNet pre-trained weights encoding the structural information and unique characteristics of the corresponding histological region. 


\textbf{Bladder Dataset --}
The skin data contains 457 cases with 6 different subtypes as classes. The subtypes include `Transitional cell carcinoma', `Papillary transitional cell carcinoma', `Cholangiocarcinoma', `Squamous cell carcinoma-NOS',  `Papillary adenocarcinoma-NOS', `Carcinoma-NOS'.

\textbf{Breast Dataset --}
The breast dataset contains 1,133 WSIs, with 23 subtypes including `Lobular carcinoma-NOS', `Infiltrating duct carcinoma-NOS', `Infiltrating duct mixed with other types of carcinoma', `Adenoid cystic carcinoma', `Infiltrating duct and lobular carcinoma', `Intraductal micropapillary carcinoma', `Large cell neuroendocrine carcinoma', `Apocrine adenocarcinoma',  `Mucinous adenocarcinoma', `Metaplastic carcinoma-NOS', `Infiltrating lobular mixed with other types of carcinoma',  'Carcinoma-NOS', `Paget disease and infiltrating duct carcinoma of breast',  `Pleomorphic carcinoma', `Papillary carcinoma-NOS',  `Phyllodes tumor-malignant', `Secretory carcinoma of breast', `Intraductal papillary adenocarcinoma with invasion', `Cribriform carcinoma-NOS', `Medullary carcinoma-NOS',  `Tubular adenocarcinoma', `Basal cell carcinoma-NOS',  `Malignant lymphoma-large B-cell-diffuse-NOS'.

\textbf{Kidney Dataset --}
the kidney contains 6 subtypes and 940 cases fro TCGA. The subtypes include `Papillary adenocarcinoma-NOS', `Clear cell adenocarcinoma-NOS', `Renal cell carcinoma-NOS', `Giant cell sarcoma', `Renal cell carcinoma-chromophobe type', `Synovial sarcoma-spindle cell'.

\textbf{Colon Dataset --}
The colon dataset has 459 cases from TCGA that contain 9 classes, `Adenocarcinoma-NOS', `Mucinous adenocarcinoma', `Adenosquamous carcinoma',  `Adenocarcinoma with neuroendocrine differentiation',  `Adenocarcinoma with mixed subtypes', `Carcinoma-NOS', `Papillary adenocarcinoma-NOS', `Dedifferentiated liposarcoma',  `Malignant lymphoma-large B-cell-diffuse-NOS'.



\begin{table*}[h]
    \centering
    \begin{tabular}{lccc}
        Experiment & Accuracy & Macro $\bar{F}1$ & Weighted $\bar{F}1$ \\
        \toprule
        Average Pooling & 0.820 & 0.448 & 0.789 \\
        Deep Binary Fisher Vector & \cellcolor{mygreen}0.852 & \cellcolor{mygreen}0.471 & \cellcolor{mygreen}0.827 \\
        Deep Sets (Max) & 0.835 & 0.473 & 0.804 \\
        Deep Sets (Mean) & 0.846 & 0.513 & 0.828 \\
        Deep Sets (Prod) & 0.824 & 0.470 & 0.799 \\
        Deep Sets (Sum) & 0.833 & 0.427 & 0.797 \\
        Deep Sparse Fisher Vector & \cellcolor{mygreen}0.859 & \cellcolor{mygreen}0.525 & \cellcolor{mygreen}0.835 \\
        Fisher Vector &  \cellcolor{mygreen}0.857 & \cellcolor{mygreen}0.510 & \cellcolor{mygreen}0.830 \\
        Focal Attention (Max) & 0.818 & 0.435 & 0.786 \\
        Focal Attention (Mean) & 0.800 & 0.392 & 0.761 \\
        Focal Attention (Sum) & 0.831 & 0.398 & 0.791 \\
        Max Pooling & 0.829 & 0.486 & 0.810 \\
        Memory Network (Max) & 0.818 & 0.413 & 0.803 \\
        Memory Network (Mean) & 0.796 & 0.329 & 0.768 \\
        Memory Network (Prod) & 0.802 & 0.450 & 0.780 \\
        Memory Network (Sum) & 0.837 & 0.362 & 0.780 \\ \hline
        Yottixel Median of Minima & 0.796 & 0.375 & 0.757 \\
        \bottomrule
    \end{tabular}
    \caption{Bladder Metrics}
    \label{tab:bladder_metrics}
\end{table*}

\begin{table*}[h]
    \centering
    \begin{tabular}{llll}
        Experiment & Accuracy & Macro $\bar{F}1$ & Weighted $\bar{F}1$ \\
        \toprule
        Average Pooling & 0.667 & 0.103 & 0.615 \\
        Deep Binary Fisher Vector & \cellcolor{mygreen}0.695 & \cellcolor{mygreen}0.125 & \cellcolor{mygreen}0.644 \\
        Deep Sets (Max) & 0.640 & 0.111 & 0.613 \\
        Deep Sets (Mean) & 0.626 & 0.095 & 0.604 \\
        Deep Sets (Prod) & 0.644 & 0.109 & 0.611 \\
        Deep Sets (Sum) & 0.636 & 0.111 & 0.613 \\
        Deep Sparse Fisher Vector & \cellcolor{mygreen}0.692 & \cellcolor{mygreen}0.132 & \cellcolor{mygreen}0.647 \\
        Fisher Vector & \cellcolor{mygreen}0.692 & \cellcolor{mygreen}0.146 & \cellcolor{mygreen}0.639 \\
        Focal Attention (Max) & 0.652 & 0.086 & 0.605 \\
        Focal Attention (Mean) & 0.647 & 0.085 & 0.599 \\
        Focal Attention (Sum) & 0.629 & 0.073 & 0.578 \\
        Max Pooling & 0.640 & 0.100 & 0.601 \\
        Memory Network (Max) & 0.588 & 0.068 & 0.545 \\
        Memory Network (Mean) & 0.622 & 0.068 & 0.556 \\
        Memory Network (Prod) & 0.621 & 0.067 & 0.526 \\
        Memory Network (Sum) & 0.592 & 0.061 & 0.472 \\
        Yottixel Median of Minima & 0.620 & 0.068 & 0.564 \\
        \bottomrule
    \end{tabular}
    \caption{Breast Metrics}
    \label{tab:breast_metrics}
\end{table*}

\begin{table*}[h]
    \centering
    \begin{tabular}{llll}
        Experiment & Accuracy & Macro $\bar{F}1$ & Weighted $\bar{F}1$ \\
        \toprule
        Average Pooling & 0.702 & 0.469 & 0.681 \\
        Deep Binary Fisher Vector & \cellcolor{mygreen}0.802 & \cellcolor{mygreen}0.590 & \cellcolor{mygreen}0.798 \\
        Deep Sets (Max) & 0.763 & 0.549 & 0.756 \\
        Deep Sets (Mean) & 0.760 & 0.549 & 0.755 \\
        Deep Sets (Prod) & 0.747 & 0.535 & 0.741 \\
        Deep Sets (Sum) & 0.764 & 0.547 & 0.756 \\
        Deep Sparse Fisher Vector & \cellcolor{mygreen}0.793 & \cellcolor{mygreen}0.586 & \cellcolor{mygreen}0.787 \\
        Fisher Vector & \cellcolor{mygreen}0.799 & \cellcolor{mygreen}0.572 & \cellcolor{mygreen}0.788 \\
        Focal Attention (Max) & 0.534 & 0.301 & 0.503 \\
        Focal Attention (Mean) & 0.568 & 0.351 & 0.544 \\
        Focal Attention (Sum) & 0.555 & 0.315 & 0.525 \\
        Max Pooling & 0.690 & 0.464 & 0.669 \\
        Memory Network (Max) & 0.644 & 0.364 & 0.589 \\
        Memory Network (Mean) & 0.650 & 0.393 & 0.628 \\
        Memory Network (Prod) & 0.639 & 0.377 & 0.596 \\
        Memory Network (Sum) & 0.492 & 0.163 & 0.331 \\
        Yottixel Median of Minima & 0.472 & 0.248 & 0.441 \\
        \bottomrule
    \end{tabular}
    \caption{Kidneys Metrics}
    \label{tab:kidneys_metrics}
\end{table*}

\begin{table*}[h]
    \centering
    \begin{tabular}{lccc}
        Experiment & Accuracy & Macro $\bar{F}1$ & Weighted $\bar{F}1$\\
        \toprule
        Average Pooling & 0.807 & 0.322 & 0.771 \\
        Deep Binary Fisher Vector & 0.809 & \cellcolor{mygreen}0.302 & 0.789 \\
        Deep Sets (Max) & \cellcolor{mygreen}0.839 & \cellcolor{mygreen}0.377 & \cellcolor{mygreen}0.795 \\
        Deep Sets (Mean) & 0.802 & 0.317 & 0.779 \\
        Deep Sets (Prod) & \cellcolor{mygreen}0.830 &\cellcolor{mygreen} 0.343 & 0.785 \\
        Deep Sets (Sum) & 0.818 & 0.297 & 0.782 \\
        Deep Sparse Fisher Vector & 0.807 & 0.281 & 0.776 \\
        Fisher Vector & 0.825 & 0.313 & \cellcolor{mygreen}0.788 \\
        Focal Attention (Max) & 0.807 & 0.332 & 0.770 \\
        Focal Attention (Mean) & 0.816 & 0.322 & 0.769 \\
        Focal Attention (Sum) & 0.825 & 0.327 & 0.776 \\
        Max Pooling & 0.825 & 0.312 & \cellcolor{mygreen}0.784 \\
        Memory Network (Max) & 0.805 & 0.285 & 0.758 \\
        Memory Network (Mean) & 0.798 & 0.301 & 0.754 \\
        Memory Network (Prod) & 0.820 & 0.291 & 0.760 \\
        Memory Network (Sum) & \cellcolor{mygreen}0.843 & 0.297 & 0.773 \\
        Yottixel Median of Minima & 0.802 & 0.298 & 0.757 \\
        \bottomrule
    \end{tabular}
    \caption{Colon Metrics}
    \label{tab:colon_metrics}
\end{table*}

\subsection{Benchmarking Scheme and Baselines}

WSI search and retrieval offers many benefits including teleconsultation, reduced workload, improved diagnostic quality, and expedited adoption and democratization of digital pathology through more efficient indexing of tissue images  \cite{tizhoosh2018artificial,janowczyk2016deep}. Considering these benefits, we are motivated to investigate the quality of WSI embeddings through the lens of $k$-Nearest Neighbors ($k$-NN) WSI search and retrieval. 

The accuracy, Macro F1 score, and weighted F1 score performance measures are used as the evaluation metrics. We benchmark multiple techniques against Yottixel's ``median of minimums'' as a non-aggregation approach \cite{kalra2020yottixel}. We test max and mean poolings, Gaussian Mixture Model-based Fisher Vector \cite{sanchez2013image}, Deep Sets \cite{zaheer2017deep}, memory networks \cite{kalra2020learning}, focal attention \cite{kalra2021pay}, and deep sparse Fisher Vector and deep binary Fisher Vector by \cite{hemati2023learning}. 

Except for the median of minimums, all methods lead to a single embedding per WSI. Further, the deep Fisher Vector provides both binary and sparse WSI embeddings which are ideal for fast and efficient WSI search. These representations are memory-efficient, ensuring that the embeddings can be economically stored and efficiently processed for subsequent retrieval and analysis tasks. To ensure the reliability of the results, a rigorous 5-fold cross-validation approach is employed. We compute the average and standard deviation from the five splits and use these as performance indicators.

\section{Results}

Tables \ref{tab:bladder_metrics}, \ref{tab:breast_metrics}, \ref{tab:kidneys_metrics} and \ref{tab:colon_metrics} show the evaluation results for bladder, breast, kidney, and colon datasets where \emph{DenseNet} was used as feature extractors. Clearly, in most cases, Deep Fisher Vector and its variations (sparse and binary) achieve the best overall performance compared with all other baselines.  Here we should emphasize sparse embeddings can reduce the memory required to store embeddings while binary WSI embeddings provide both memory efficiency and extremely fast search speed by employing Hamming distance. 
As a sample case, we also measured the time required to conduct the colon dataset search experiment for both Deep Sparse and Binary Fisher Vectors and also the median of minimums. For the Sparse Fisher Vector, Euclidean distance is used while for the Binary Fisher Vector Hamming distance is employed. The results for this experiment are presented in Table \ref{tab:search_time_colon}. Clearly, the Deep Binary Fisher Vector achieves the fastest search speed. Given that for the Sparse Fisher Vector the Euclidean distance is used, its search speed is the same as other methods that output one real-valued embedding (with the same dimension) per WSI.

\begin{table}[htb]
    \centering
    \begin{tabular}{l|c|c|c}
        \hline
        Dimensions & Sparse & Binary & Yottixel \\
        \hline
        30,000 & 1,659,800 ms & 690,026 ms & \\
        3,000  & 1,491,833 ms & 601,359 ms & 1,007,515 ms\\
        300    & 1,815,617 ms & 475,188 ms \\
        \hline 
    \end{tabular}
    \caption{Time required for different dimensions of Fisher vector for Sparse and Binary methods for the Colon dataset compared to Yottixel}
    \label{tab:search_time_colon}
\end{table}

\subsection{Ablation study}

To study the effect of the gradient sparsity and quantization losses for Deep Sparse and Binary Fisher vector embeddings respectively, we performed an ablation study on $\alpha$ for skin and lung datasets. To this end, we train the VAE for different values of $\alpha$,  while keeping all other hyperparameters the same. The results for breast and colon datasets are presented in Tables \ref{tab:breast_ablation_alpha_actual} and \ref{tab:colon_ablation_alpha}. Clearly, including the regularization loss further improves the quality of embeddings while increasing memory and speed efficiency by introducing binary nature and sparsity characteristics for WSI embedding.

\begin{table*}[h]
    \centering
    \begin{tabular}{l|c c c|c c c}
        \hline
        $\alpha$ & \multicolumn{3}{c|}{Breast (Sparse)} & \multicolumn{3}{c}{Breast (Binary)} \\
        & Accuracy & Macro $\bar{F}1$ & Weighted $\bar{F}1$ & Accuracy & Macro $\bar{F}1$ & Weighted $\bar{F}1$ \\
        \hline
        0.0     & 0.690 & 0.151 & 0.651 & 0.685 & 0.126 & 0.640 \\
        0.1     & 0.693 & 0.149 & 0.650 & 0.703 & 0.131 & 0.653 \\
        0.01    & 0.698 & 0.140 & 0.651 & 0.697 & 0.123 & 0.646 \\
        0.001   & 0.702 & 0.145 & 0.660 & 0.684 & 0.115 & 0.636 \\
        0.0001  & 0.688 & 0.137 & 0.646 & 0.691 & 0.114 & 0.637 \\
        0.00001 & 0.706 & 0.144 & 0.663 & 0.680 & 0.116 & 0.634 \\
        \hline
    \end{tabular}
    \caption{Results for Sparse and Binary Fisher vector methods for Breast dataset with different values of $\alpha$, including Accuracy, Macro $\bar{F}1$, and Weighted $\bar{F}1$}
    \label{tab:breast_ablation_alpha_actual}
\end{table*}

\begin{table*}[h]
    \centering
    \begin{tabular}{l|c c c|c c c}
        \hline
        $\alpha$ & \multicolumn{3}{c|}{Colon (Sparse)} & \multicolumn{3}{c}{Colon (Binary)} \\
        & Accuracy & Macro $\bar{F}1$ & Weighted $\bar{F}1$ & Accuracy & Macro $\bar{F}1$ & Weighted $\bar{F}1$ \\
        \hline
        0.0     & 0.825 & 0.358 & 0.789 & 0.807 & 0.274 & 0.778 \\
        0.1     & 0.820 & 0.339 & 0.782 & 0.807 & 0.307 & 0.785 \\
        0.01    & 0.820 & 0.340 & 0.782 & 0.811 & 0.277 & 0.783 \\
        0.001   & 0.814 & 0.330 & 0.776 & 0.818 & 0.288 & 0.790 \\
        0.0001  & 0.816 & 0.316 & 0.778 & 0.820 & 0.312 & 0.786 \\
        0.00001 & 0.816 & 0.334 & 0.780 & 0.805 & 0.301 & 0.779 \\
        \hline
    \end{tabular}
    \caption{Results for Sparse and Binary Fisher vector methods for Colon dataset with different values of $\alpha$, including Accuracy, Macro $\bar{F}1$, and Weighted $\bar{F}1$}
    \label{tab:colon_ablation_alpha}
\end{table*}






\section{Conclusion}

In this paper, we evaluated multiple WSI representation learning techniques including simple average or max pooling operations, deep sets, memory networks, focal attention, Gaussian Mixture Model (GMM) Fisher Vector, and deep sparse and binary Fisher Vector embeddings in the WSI search task for different datasets and two different deep models. Our evaluations show that the deep sparse and binary Fisher Vector may achieve the best overall performance compared with multiple set aggregation schemes. This result suggests that the Fisher Vector method has a robust ability to capture relevant patterns and features within the data, regardless of the specific characteristics of each dataset. This is an important indication of the versatility and generalizability of the Fisher Vector approach. Apart from their superior WSI search performance,  deep sparse and binary Fisher Vectors also offer additional advantages that are suitable for fast and efficient search and retrieval. More precisely, the sparsity in deep sparse Fisher Vector embeddings makes them occupy less storage for indexing WSIs, a crucial factor for the wide adoption and also democratization of digital pathology. Also, the deep binary Fisher Vector is significantly faster than any real-valued counterpart as the calculation of Hamming distance between binary embeddings can be calculated using the XOR gate at the CPU level which is extremely faster than Euclidean distance.

{\small
\bibliographystyle{unsrt}
\bibliography{IEEEexample}
}

\end{document}